\documentclass[12pt]{article}  
\usepackage{setspace}
\usepackage{abstract}
\usepackage{tabularx}
\usepackage{enumerate}
\usepackage[authoryear]{natbib}
\usepackage{epsfig, latexsym, multirow}
\usepackage{bbold}
\pdfoutput=1
\usepackage[utf8]{inputenc}
\usepackage[T1]{fontenc}
\usepackage{float}
\usepackage{authblk}
\usepackage[normalem]{ulem}
\usepackage{amsmath}
\usepackage{amsthm}
\usepackage{booktabs}
\usepackage{mathtools}
\usepackage{fullpage}
\usepackage{subcaption}
\usepackage{wrapfig}
\usepackage{caption}

\newcommand{\comment}[1]{}
\usepackage{graphicx,scalerel}
\usepackage{parskip}
\usepackage{ragged2e}

\usepackage[table,xcdraw]{xcolor}
\usepackage{amssymb}
\usepackage[colorlinks = true,
linkcolor = black,
urlcolor  = black,
citecolor = teal]{hyperref}

\date{}

\begin{document}
	\normalem
	
	\title{Nonparametric Estimation of the Underlying Distribution of Binned Continuous Data}
	
	\author{
		Ejike R. Ugba\thanks{To whom correspondence should be addressed:
			\texttt{ugbae@hsu-hh.de}} \thanks{Department of Mathematics and Statistics, Helmut Schmidt University, Hamburg, Germany} \ \& Jan Gertheiss$^\dagger$ 
		
	}
	
	\maketitle
	
	\begin{center}
		\textbf{Abstract}
	\end{center}
	
	Estimating cumulative distribution functions (CDF) and probability density functions (PDF) is a fundamental practice in applied statistics. However, challenges often arise when dealing with data arranged in grouped intervals. This paper discusses a suitable and highly flexible non-parametric density estimation approach for binned distributions, based on cubic monotonicity-preserving splines—known as cubic spline interpolation. Results from simulation studies demonstrate that this approach outperforms many widely used heuristic methods. Additionally, the application of this method to a dataset of train delays in Germany and micro census data on distance and travel time to work yields both meaningful but also some questionable results.
	
	\vspace{1cm}
	
	
	\newpage

	%----------------------------------------------------
	\section{Introduction}\label{intro}
	%----------------------------------------------------
	Nonparametric density estimation has gained much attention in diverse fields of empirical research. Unlike its parametric counterpart with many restrictive assumptions, this approach is independent of distributional assumptions. The literature is inundated with many valuable materials on the nonparametric estimation of density functions. Some benchmark references include \cite{silverman1986a}, \cite{eubank1988a}, \cite{loader1999a}, and \cite{wasserman2006a}. The cardinal point of most nonparametric estimations hinges on the estimation of probability density functions (PDFs) and cumulative distribution functions (CDFs), considered crucial cornerstones of statistical estimation and inference. However, when information on a supposedly continuous attribute of interest is captured in groups of intervals rather than the original measurement scale, estimating the CDFs and PDFs of the underlying continuous variable is often very problematic. A common situation in practice is in surveys where data on income, for instance, is only available in terms of intervals or where an age bracket is used instead of capturing the actual age of respondents. Although some mathematical heuristics may be employed to obtain some descriptive statistics of interest, such, of course, fail to tell the full story regarding the true distribution being estimated.
	
	In what follows, we assume that the said grouped data $z$ in terms of intervals arises from an underlying continuous variable $y^*$ such that      
	\begin{equation}\label{threshold}
		z = j, \; \text{if} \  \tau_{j-1} < y^* \leq \tau_{j},
	\end{equation}
	with $j = 1,\dots, r$ and the thresholds $\tau_{j}$ being known. We could use $z$ to estimate the entire distribution of $y^*$, given that we, at least, know the empirical cumulative distribution function (eCDF) of $y^*$ at $\tau_{j}$, $j = 1,\dots, r-1$. In the simplest case, we can just use linear interpolation to estimate the entire cdf. This is exactly what is accomplished by drawing a relative cumulative frequency graph. However, linear interpolation may induce some systematic bias because it implicitly assumes that the underlying data is uniformly distributed within each interval defined by the cutpoints. This also means that the underlying pdf is piecewise constant with potential jumps at the cutpoints, a rather unrealistic assumption in many applications.

	Alternatively, one may consider (monotone) splines of higher order, such as cubic splines, to obtain a smoother and differentiable function. An example of this approach is found in a recent paper by \cite{vhippel2017a}, where a similar technique was applied when analyzing binned income data of US counties. Our study will rigorously evaluate the performance of monotone spline interpolated cumulative distribution functions (CDFs). This evaluation will be conducted through an extensive simulation study, encompassing an underlying CDF with multiple distributional shapes and support.
	
	The subsequent sections of this study are organized as follows: A concise overview of the related literature is presented in Section~\ref{lit_review}. We delve into the heuristic approach and the kernel/eCDF estimation method in Sections~\ref{heuristic} and~\ref{ecdf}, respectively. This is followed by an in-depth presentation of the interpolated CDF in Section~\ref{spline}. In Section~\ref{sim}, we conduct a comprehensive simulation study to assess the performance of the monotone spline interpolated CDF. Real data examples are then explored in Section~\ref{realDt}. Finally, Section~\ref{disc} offers a conclusive discussion of our findings.

	%-------------------------------------------------------------------------------
	\section{Previous Work}\label{lit_review}
	%-------------------------------------------------------------------------------
	A readily available nonparametric approach to density estimation is the so-called kernel density estimation, the main ideas of which can be traced back to works such as \cite{Rosenblatt1956}, \cite{Parzen1962}, \cite{nadaraya1964a}, \cite{yamato1973a}, and \cite{azzalini1981a}. These early contributions employed kernel functions to estimate the distribution of univariate, independent, and identically distributed (iid) random variables. \cite{liu2008a} investigated the asymptotic properties of the integrated smooth kernel estimator for multivariate and weakly dependent data. In contrast, \cite{funke2016a} explored estimating non-negative valued random variables via convolution power kernels. Despite these advancements, selecting an appropriate bandwidth for kernel-type estimation remains an issue in many empirical applications.
	
	An underexplored avenue of nonparametric density estimation involves using polynomial curves and spline functions to estimate the cumulative distribution functions (CDFs) and probability density functions (PDFs) of measurement attributes. \cite{kim1996a}, for instance, discusses the use of Bezier curves, a variant of X-splines, for nonparametric density estimation. \cite{kim1999s} demonstrated that such an application of Bezier curves competes favorably with results obtained from kernel-based estimators and shares the same asymptotic properties. Bezier curves were also employed by \cite{kimBiz2003a} for smoothing the Kaplan-Meier estimator and by \cite{bae2005a} for smoothing a bivariate Kaplan-Meier estimator. \cite{Kimc2012a} discussed selection techniques for Bezier points, while \cite{w2014a} discussed the nonparametric estimation of the distribution function using the Bezier curve. However, the subjectivity in determining the shape parameter and the number of adequate Bezier points in various situations remains a concern.
	
	Monotone spline estimators were employed by \cite{l2009a} to estimate the CDFs of random variables in simulation and real data studies. \cite{huang2016a} applied spline and kernel estimators to estimate the interval-aggregated data's PDF. Furthermore, \cite{charles2010a} used cubic monotone control smoothing splines to estimate finite interval-defined CDFs. \cite{Rizzi2015} demonstrated how binned data can be ungrouped via a penalized composite link model, assuming the original distribution meets a specific smoothness criterion. \cite{Jargowsky2018} proposed a mean-constrained integration over brackets (MCIB) approach to improve existing binned data estimation methods. In a recent paper, \cite{ghosh2018a} proposed a method for estimating the latent density of ordinal data using a combination of Bernstein Polynomials and information on latent distribution provided by cut-points. Their estimation approach is based on the Anderson-Darling statistic rather than the widely used maximum likelihood estimation and yields good results. However, such an approach is plagued with a loss of efficiency arising from its inability to constrict the shape of the latent density to unimodal when such a situation arises. Some other handful attempts have also been made in the past towards estimating the distribution of binned data using both parametric and nonparametric approaches, which in many instances fail to perform adequately given the discontinuous structure of such data. As hinted in Section~\ref{intro}, the method used by \cite{vhippel2017a} seems promising and will be explored further in this study.

	%-------------------------------------------------------------------------------
	\section{Heuristic Estimator}\label{heuristic}
	%-------------------------------------------------------------------------------
	
	A simple and widely used method of binned data estimation is the midpoint estimator (ME), which assigns cases to all class midpoints, enabling easy computation of some sample or population characteristics of interest. Consider the threshold model in~\eqref{threshold}, where the distinct bin frequencies, $\eta_{1}, \eta_{2}, \dots, \eta_{r}$ (as shown in Table~\ref{freqDistribution}) are subdivisions of the total observation $n$.  
	Assuming $\tau_{j-1}$ and $\tau_{j}$ are finite and known, a straightforward estimation approach is to assign corresponding cases to the respective bin midpoints with the midpoint estimator 
	\begin{equation}
		\text{ME}_j = (\tau_{j-1}+ \tau_{j})/2, \quad j = 1,\dots,r, 
	\end{equation}
	\cite[see, e.g.,][]{heitjan1989a} and then compute some basic descriptive statistics. The bin consistency property of ME and its other desirable properties \cite[see][]{hippel2015a} makes it widely acceptable for most grouped data analysis. A different dimension to it is with open-ended grouped data where the last class limit, $\tau_{r}$, is considered finite but unknown (e.g., in most grouped income data), a common practice is to assume the particular bin in question follows a Pareto distribution with shape parameter $\gamma>1$ \citep{henson1967a}, and with its midpoint given by $\mu_{r}=\tau_{r-1}\gamma/(\gamma-1)$, the so-called Pareto midpoint estimator (PME) \cite[see][]{henson1967a}. \cite{hippel2015a} consider the harmonic rather than the arithmetic mean of the Pareto distribution a more robust choice, given instead the following robust Pareto mean estimator (RPME): $\mu'_{r}=\tau_{r-1}(1+1/\gamma)$, where the estimate of $\gamma$ is given by
	\begin{equation}
		\hat\gamma=\frac{\ln \{(\eta_{r-1} + \eta_{r})/\eta_{r}\}}{\ln(\tau_{r-1} / \tau_{r-2})}   % 
	\end{equation}
	
	\begin{table}[bht]\centering
		\captionsetup{font=scriptsize}
		\caption{\label{freqDistribution}Frequency distribution of a binned continuous variable ($y^*$) with $L_{b}$, $U_{b}$ and $\eta$ 
			the lower and upper class limits, and the frequencies, respectively.}
		\begin{tabular}{lc}
			\toprule[0.1 em]
			$L_{b}$ 	\quad  \hspace{45pt}        \quad   $U_{b}$         & \quad \quad $\eta$\\
			\hline
			$\tau_{0}   \quad     \ \ \ \leq y^* < \quad   \tau_{1}$        & \quad \quad $\eta_{1}$ \\
			$\tau_{1}   \quad     \ \ \ \leq y^* < \quad   \tau_{2}$        & \quad \quad $\eta_{2}$ \\ 
			$\vdots$ 	       	                                            & \quad \quad $\vdots$ \\
			$\tau_{r-2} \quad           \leq y^* < \quad   \tau_{r-1} $     & \quad \quad $\eta_{r-1}$ \\
			$\tau_{r-1} \quad           \leq y^* < \quad   \tau_{r}$        & \quad \quad $\eta_{r}$ \\
			\bottomrule[0.05 em]
		\end{tabular}
	\end{table}

	However, while such an approach does provide a quick remedy to the problem at hand, it goes without saying that such an imposed parametric assumption on a single data point could yield unreliable population estimates and misleading conclusions. Consequently, a few parametric approaches have been proposed. For instance, the generalized beta (GB) family has been proposed \citep{mcdonald2008a} and applied \cite[see,][]{hippel2015a} in several binned data arising from large-scale survey studies. To improve the utility of the GB family, a multimodel GB estimator (MGBE) was further developed in \cite{hippel2015a}. However, besides not yielding correct fit to data in most empirical applications, the GB family dependence on iterative methods slows things down, making it difficult to use in large-scale studies \citep{vhippel2017a}. 
	
	In addition to the familiar measures of central tendency and spread, in most instances, it is usually of interest to obtain different fractiles of binned data, including quartiles, deciles, percentiles, and n-tiles in general. Thus, with binned data, such quantities are generally obtained either with linear CDF interpolation of observed classes (a relative frequency distribution graph) or the widely used interpolation formula, both yielding the same result. For instance, the quantiles of a given binned data could be obtained as follows 
	\begin{equation}
		Q(q) = \tau_L +  \frac{nq - C}{\eta}W, \quad  q \in (0,1),
	\end{equation}
	where $q$ is the desired quantile (a value between 0 and 1) and $n$ is the total number of observations. Assuming $\xi$ denotes the class containing $q$, $\tau_L$ is the lower boundary of  $\xi$, $C$ is the cumulative frequency of the class preceding $\xi$, $\eta$ is the frequency of $\xi$, and $W$ is the width of $\xi$. The following special quantile: $Q_{.25}$, $Q_{.50}$ and $Q_{.75}$ respectively denote the first quartile $(Q_1)$, the median $(Q_2)$, and the third quartile $(Q_3)$ of a given distribution, with the interquartile range given by 
	
	\begin{equation}
		IQR = Q_3 - Q_1.
	\end{equation}
	
	These and similar measures are often employed empirically to study several population characteristics of interest. Our goal in this study is to examine an alternative non-parametric CDF and density estimation approach with which such measures could be optimally estimated and to compare the same with the status quo.
	
	In the next section, we present a formal definition of the CDF, how it relates to the PDF and other advanced approaches to their estimations.

	%--------------------------------------
	\section{Kernel and eCDF Estimators}\label{ecdf}
	%-------------------------------------- 
	Assuming $X_{1}, X_{2}, \dots X_{n}$ are independent identically distributed real-valued random variables with a continuous PDF $f(\tau)$, and $F_{X}(\tau) = P(X \leq \tau)$ denotes the cumulative distribution function (CDF) of the same, where the right-hand side represents the probability of $X$ taking on a value less than or equal to $\tau$. $F_{X}(\tau)$ represents a right-continuous monotonically increasing function \citep{park2018a} that uniquely describes $f(\tau)$. For a continuous $X$, the following relationship between $F_{X}(\tau)$ and $f(\tau)$ applies:
	
	\begin{equation}
		F_{X}(\tau) = \int_{-\infty}^{\tau} f_{X}(t) \,dt.
	\end{equation}
	
	This relationship implies that it is possible to obtain $F_{X}(\tau)$ from $f(\tau)$ if known and vice versa. As earlier hinted in Section~\ref{lit_review}, several non-parametric approaches to CDF and density estimations already exist in the literature. A typical and common one is the so-called kernel density estimator \citep{Rosenblatt1956, Parzen1962}. In an ideal continuous case, where $X_{1}, X_{2}, \dots X_{n}$ are supposed to be observed, the kernel estimator of the CDF is expressed as follows,
	
	\begin{equation}\label{kernel}
		\hat{F_{h}}(\tau) =  \frac{1}{nh}  \sum_{i=1}^{n} \mathbb{\kappa} \left(\frac{\tau-X_{i}}{h}\right),
	\end{equation}

	where $\kappa(x) = \int_{-\infty}^{x} K(t) \, dt$ is an integral over a kernel function $K$, with the bandwidth denoted by $h$. While the choice of $K$ is of secondary importance, that of $h$ plays a crucial role in the shape of the estimator given in (\ref{kernel}). Two widely used selection approaches for $h$ include cross-validation and plug-in methods \citep{Jones1996, Loade1999, Henderson2015}. While the first method offers a data-driven solution to the estimation problem, the second approach seeks to minimize the asymptotic mean integrated squared error while substituting the unknown density in the optimization with a pilot estimate. These notwithstanding, the selection of $h$ and extension to non-continuous data still pose serious concerns in practice. For instance, trying to apply the second approach to the group midpoints of the grouped data could result in unreliable estimates since too much probability mass is falsely allocated to the midpoints and too little to the unobserved $X_i$ \citep{Scott1985, Wang2013}. Unfortunately, choosing a large $h$ to circumvent such problems incurs a further loss of information about the underlying true distribution \citep{Reyes2019}, thus offering no helpful solution to the problem. Furthermore, \cite{Minoiu2014} analyzed the performance of kernel density methods applied to grouped data to estimate poverty using Monte Carlo simulations and household surveys, with findings indicating that such technique gives rise to biases in poverty estimates.	
	
	A different frequently used estimator of $F_{X}(\tau)$ is the empirical cumulative distribution function (eCDF) denoted by $\hat{F_{n}}(\tau)$. Given $n$ number of observations, $\hat{F_{n}}(\tau)$ assigns a relative frequency (i.e., estimated probability) of $1/n$ to an ordered sequence of the observed data, resulting in a step function that increases by $1/n$ at each datum. This may be formally expressed as follows:
	
	\begin{equation}
		\hat{F_{n}}(\tau) = \frac{1}{n}  \sum_{j=1}^{n} \mathbb{1} \{X_j \leq \tau  \}  \quad \forall \tau \in \rm I\!R 
	\end{equation}
	
	\begin{figure}[t]\centering
		\includegraphics[width=120mm]{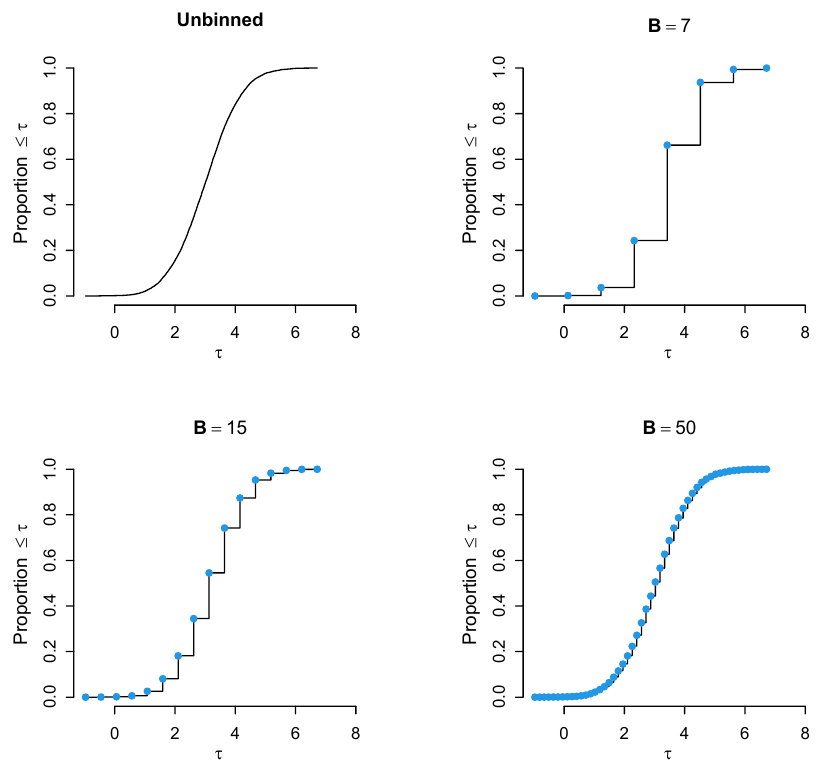}
		\captionsetup{font=scriptsize}
		\caption{\label{ecdfs} The empirical cumulative distribution function (eCDF, unbinned) of an iid normally distributed random sample $(n=10000)$, and the eCDF after the observations are binned, considering a small, medium, and large number of bins $B$.}
	\end{figure}
	
	where $\mathbb{1}\{\cdot\}$ is the indicator function, $\lim_{\tau\to -\infty} F_{X}(\tau) = 0$ and $\lim_{\tau\to\infty} F_{X}(\tau) = 1$. However, similar to the previously discussed estimator, estimating an underlying CDF from eCDF makes absolute sense with ungrouped continuous random variables, given a sufficiently large number of observations. Grouped data requires having a large number of bins to make any meaningful approximation of an underlying CDF. One has that. 
	
	\begin{equation}
		F_{X}(\tau) = \lim_{B\to \infty} F_{X_B}(\tau),
	\end{equation}
	
	where $F_{X_B}(\tau)$ is the eCDF of a binned data and $B$ the number of bins. In practice, a relatively small number of bins ($B$ < 10 or a bit more) is typically used to capture various population characteristics of interest, such that eCDFs obtained under such settings, irrespective of the total number of frequencies, are mere step functions with large jumps that are not differentiable. This is further illustrated in Figure~\ref{ecdfs}, showing the eCDF of an unbinned random sample from the normal distribution and the binned version of the same with $B \in \{7, 15, 50\}$. As observed, with a large number of bins, $F_{X_B}(\tau)$ approximates $F_{X}(\tau)$. However, as a large $B$ is practically unattainable for single or multiple study characteristics, one at least needs a smooth differentiable function, irrespective of $B$, with which the underlying distribution could be approximated. Thus, we will next consider interpolating the CDFs with splines of higher order.

	%-------------------------------------------------------------------------------
	\section{Monotone Spline Interpolated CDF}\label{spline}
\begin{figure}[t]
		\centering
		\includegraphics[width=150mm]{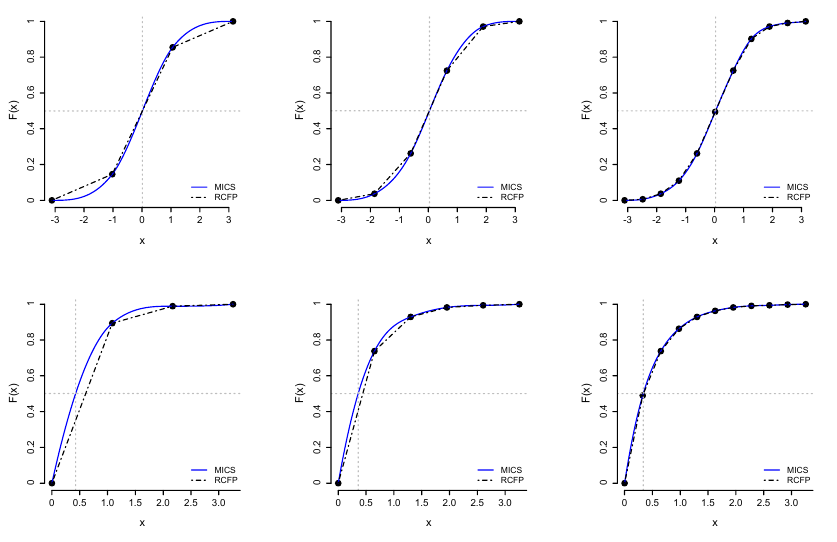}
		\captionsetup{font=scriptsize}
		\caption{\label{mics-rcfp} Estimated CDF of the standard normal (upper row) and the gamma (lower row) distributions for $r = \{3, 5 \mbox{ and } 10\}$ using monotone interpolating cubic splines (MICS) and relative cumulative frequency plots (RCFP). Dashed grey horizontal and vertical lines indicate the second quartile on MICS compared to RCFP.}
	\end{figure}

	%-------------------------------------------------------------------------------
	A basic and prevalent form of CDF interpolation for binned data is the relative cumulative frequency plot (RCFP), which maps the cutpoints $\tau_j$ of binned data to the realized cumulative frequencies, connecting them with straight lines \cite[see, e.g.,][]{vhippel2017a}. However, RCFP produces a function with rough edges at the respective connecting nodes and, as such, does not represent a smooth function. At best, it offers a crude estimation of an underlying CDF. A robust and more accurate estimate of $F_{X}(\tau)$ from binned data could be achieved using appropriate monotone interpolating cubic splines (MICS) or similar functions \citep{vhippel2017a}. While a more elaborate simulation study follows in the next section, to buttress the point a bit more, Figure~\ref{mics-rcfp} shows a typical estimated CDF of the standard normal (upper row) and the gamma (lower row) distributions for $r = \{3, 5 \mbox{ and } 10\}$ using MICS and RCFP. While MICS provides a smooth differentiable curve that passes through the nodes, RCFP connects the points with straight lines that are somewhat discontinuous at the nodes, even though both shapes seem to align together given a large $r$. Moreover, the indicated median (dashed grey lines) shows how both lines coincide at this point given the normal distribution but differ for the gamma distribution. 
	
	A whole lot of literature exists on interpolation approaches, with some benchmark references being papers by \cite{fritschButla1980a}, \cite{fritschCarl1980a}, \cite{lehmann1999a} and \cite{hagan2006a}.
	
	Consider a mesh of data points $\{\tau_{1}, \tau_{2}, \dots, \tau_{r}\}$ and corresponding values $\{F_{1}, F_{2}, \dots, F_{r}\}$ for a generic but unknown function $F(\tau)$. Given the following definitions of the mesh sizes and gradients:
	
	\begin{equation}
		h_j = \tau_{j+1} - \tau_j
	\end{equation}
	
	\begin{equation}
		m_j = \frac{F_{j+1} - F_j}{h_j},
	\end{equation}
	
	with $j = 1,\dots, r-1$, the cubic interpolating spline function of (\ref{threshold}) could be defined by piecewise cubic polynomials that pass through consecutive points and may be expressed as follows:     
	
	\begin{equation} 
		S(\tau) = a_j + b_j(\tau - \tau_j) + c_j(\tau - \tau_j)^2 + d_j(\tau - \tau_j)^{3},
	\end{equation}
	
	where $\tau \in [\tau_j, \tau_{j+1}]$ and  $j = 1,\dots, r-1$, $a_j = F(\tau_j) \equiv F_j$, $b_j = F'(\tau_j)$, and the other two parameters defined as follows \citep[see,][]{hagan2006a}:
	
	\begin{equation}
		c_j =  \frac{3m_j - b_{j+1} - 2b_j}{h_j}
	\end{equation}
	
	\begin{equation}
		d_j = \frac{b_{j+1} + b_j - 2m_j}{h_j^{2}}
	\end{equation}
		
	The cubic interpolating spline function has the following boundary conditions: $S'(\tau_{0})=0$ and $S'(\tau_{r})=1$, where the pair $(\tau_{0}, \tau_{r})$ correspond to the minimum and maximum values of a given distribution, which can also be $-\infty$ and $\infty$, respectively.
	
	As previously noted, $F_{X}(\tau)$, which we seek to estimate, is a right-continuous monotone increasing function and, as such, would require being estimated with a corresponding monotone increasing function. This ensures that the geometry of the underlying curve is preserved. 
	
	An excellent example is Hyman's filter of the Fritsch-Butland algorithm \citep{hyman1983a, fritschButla1980a, butland1980a, fritschCarl1980a}. It offers a monotonicity-preserving algorithm that could be implemented in principle on almost any cubic-spline-like interpolation method. Hyman (1983) first defines the following.
	
	\begin{equation}
		b_1 = \frac{(2h_1 + h_2)m_{1} - h_{1} m_{2}}{h_{1} + h_2},
	\end{equation}
	
	\begin{equation}
		b_r = \frac{(2h_{r-1} + h_{r-2})m_{r-1} - h_{r-1} m_{r-2}}{h_{r-1} + h_{r-2}},
	\end{equation}
	
	followed by the popular Hyman's monotonicity constraint which ensures that no spurious extrema are introduced in the interpolated function. The constrain is given as follows \citep{hyman1983a}:

	\begin{equation}\label{hyman}
		b_j =
		\left\{ 
		\begin{array}{ c l }
			\mbox{min}(\mbox{max}(0, b_j),3\mbox{min}(m_{j-1},m_j)) & \quad \textrm{if } m_{j-1},m_j > 0 \\
			\mbox{max}(\mbox{min}(0, b_j),3\mbox{max}(m_{j-1},m_j)) & \quad \textrm{if } m_{j-1},m_j < 0
		\end{array},
		\right.
	\end{equation}
	
	where the former choice is made when the curve is increasing and the latter when the curve is decreasing. However, only the first constraint applies here since the current study focuses on an increasing $F$. See \cite{hagan2006a} for further discussion of the properties of (\ref{hyman}) and alternative approaches. 
		
	Thus, given a set of binned data as illustrated in Table~\ref{freqDistribution}, where $\tau_{0},\tau_{1},\dots,\tau_{r}$, are the thresholds (including minimum and maximum) with which the continuous attribute was discretized, and $(0,c_{1},\dots,c_{j-1}, 1)$ the corresponding cumulative relative frequencies, the underlying CDF could be mirrored by monotonicity-preserving spline interpolation of these $r$ pairs of points \citep[see, e.g.,][]{vhippel2017a}. Obtaining the underlying CDF of the binned data this way also makes it possible to realize the underlying PDF, considering that the derivative of the spline interpolator is an estimator of the density \citep{lii1975a}.

	%-------------------------------------------------------------- 
	\section{Simulation Study}\label{sim}
	%--------------------------------------------------------------
	The MICS approach was examined in a simulation study against results from the true theoretical distribution and compared to related approaches. Four distinct underlying distributions (shapes) were considered with a relatively small and large number of observations ($n = 100 \mbox{ and } 1000$). These include the normal: $y^*_{i1} \sim \mathcal{N}(3,\,1)$, gamma: $y^*_{i2} \sim \mathcal{G}(1,\,2)$, generalized extreme value: $y^*_{i3} \sim \mathcal{G}v(1,\,2, \,0)$ and triangular: $y^*_{i4} \sim \mathcal{T}(0,\,1, \,0.5)$ distributions. The finite ranges of observations $(\tau_{0}, \tau_{r})$ considered for the various distributions are respectively $(0,10)$, $(0,8)$, $(-4,35)$ and $(0,1)$.	
	
	The generated continuous distributions were discretized using $r=6$ equidistant cut-points and subsequently compressed to frequency tables, yielding a binned dataset,  $y^b_{ik}$, $k = 1, \dots 4$, with $r+1$ classes. The entire process was replicated a thousand times, resulting in a thousand pairs of datasets $\{ (y^*_{il}, y^b_{ik}), \ i = 1, \dots, n, \ k = 1, \dots, 4 \}$ for the four specified distributions.
	
	With the datasets generated, we proceeded with the estimation of the CDFs and PDFs of the individual distributions (i) directly from the continuous sampling distribution $y^*_{ik}$ and (ii) from the binned data $y^b_{ik}$, using the MICS, heuristic, eCDF, and kernel estimation approaches. These were subsequently compared to the theoretical distributions obtained with prior specified parameters. To simplify things, we shall pay particular attention to estimating quantiles, means, and standard deviations from the various distributions. The estimated quantiles with each method include .25, .50 and .75. In other words, quantiles obtained from the continuous sampling distributions and those from the binned data are compared with the true theoretical quantiles, the same with means and standard deviations. 
		
	\begin{figure}[!htb]\centering
		\includegraphics[width=168mm]{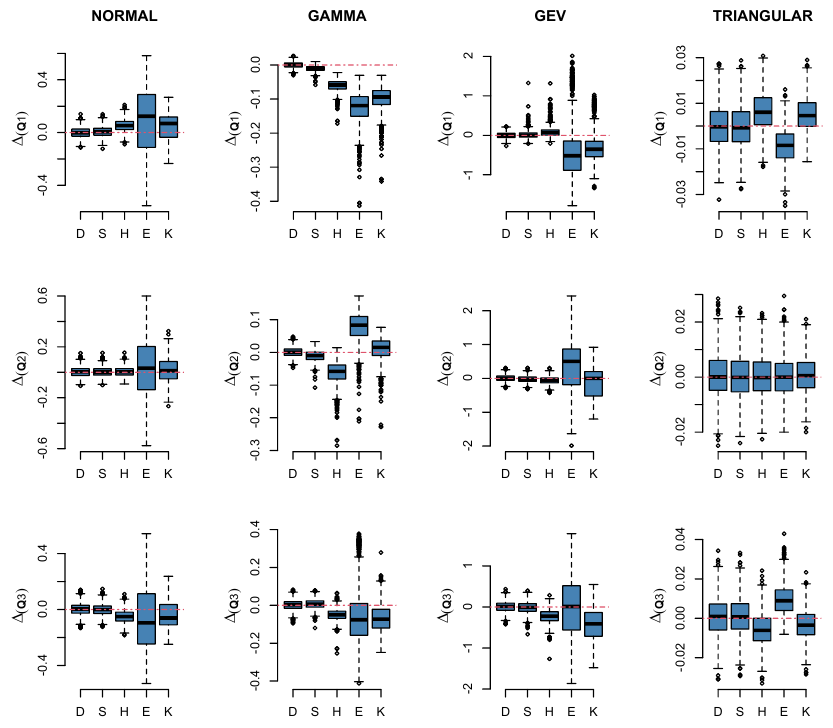}
		\captionsetup{font=scriptsize}
		\caption{\label{quantile_1000} Pairwise differences of quantile estimates $(Q_1, Q_2, Q_3)$ of binned data generated from an underlying continuous data with sample size $n=1000$. D denotes estimates from the sampling distribution, and S, H, E, and K are estimates from the spline, heuristic, eCDF, and kernel estimators, respectively. Dotted horizontal red lines represent the mark of over and underestimation of the underlying measure.}
	\end{figure}
	
	\begin{figure}[htb]\centering
		\includegraphics[width=168mm]{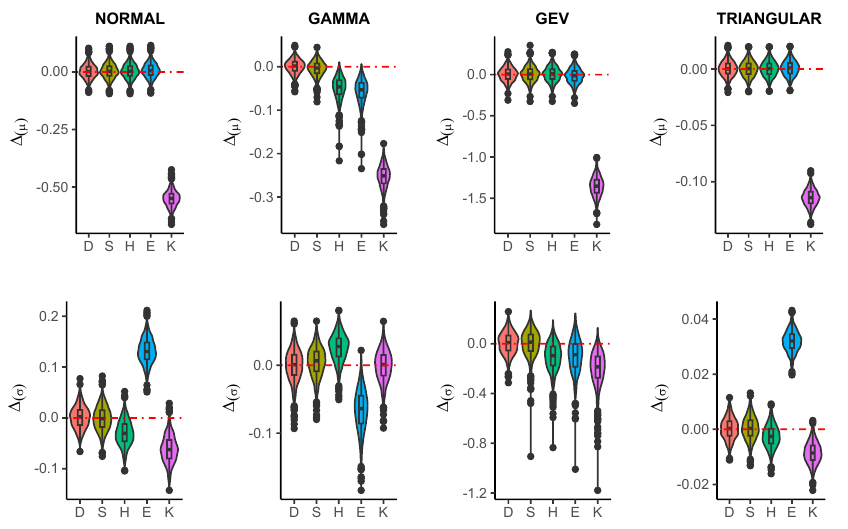}
		\captionsetup{font=scriptsize}
		\caption{\label{mean_SD_1000} Pairwise differences of mean $\Delta(\mu)$ and standard deviation $\Delta(\sigma)$ estimates of binned data generated from an underlying continuous data with sample size $n=1000$. D denotes estimates from the sampling distribution, and S, H, E, and K are estimates from the spline, heuristic, eCDF, and kernel estimators, respectively. Dotted horizontal red lines represent the mark of over and underestimation of the underlying measure.}
	\end{figure}

	For easy presentation of the results, the pairwise differences of the estimated population quantiles $\Delta(Q)$, means $\Delta(\mu)$, and standard deviations $\Delta(\sigma)$, with reference to the true theoretical values, were obtained from a large sample size $(n=1000)$ and a relatively small sample size $(n=100)$. Thus, assuming $Q_{q}$, $\mu$ and $\sigma$ respectively denote the theoretical quantiles, means, and standard deviations; $\hat{Q}_{q}$, $\hat{\mu}$, and $\hat{\sigma}$ respectively denote the estimated quantiles, means, and standard deviations from the sampling distribution or from the binned data, the paired differences are obtained as follows,

	\begin{equation}
		\Delta(Q_q) = Q_{q} -  \hat{Q}_{q}, \quad q \in (.25, \ .50, \ .75)
	\end{equation}
	
	\begin{equation}
		\Delta(\mu) = \mu - \hat{\mu}
	\end{equation}
	
	\begin{equation}
		\Delta(\sigma) = \sigma - \hat{\sigma},
	\end{equation}
	
	\begin{figure}[!htb]\centering
		\includegraphics[width=168mm]{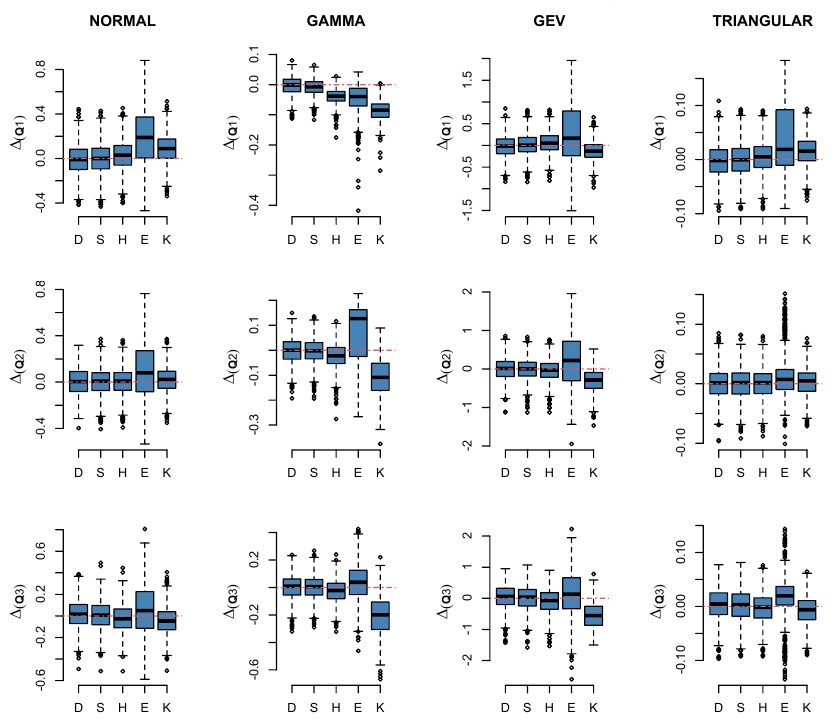}
		\captionsetup{font=scriptsize}
		\caption{\label{quantile_100} Pairwise differences of quantile estimates $(Q_1, Q_2, Q_3)$ of binned data generated from an underlying continuous data with sample size $n=100$. D denotes estimates from the sampling distribution, and S, H, E, and K are estimates from the spline, heuristic, eCDF, and kernel estimators, respectively. Dotted horizontal red lines represent the mark of over and underestimation of the underlying measure.}
	\end{figure}
	
	\begin{figure}[htb]\centering
		\includegraphics[width=168mm]{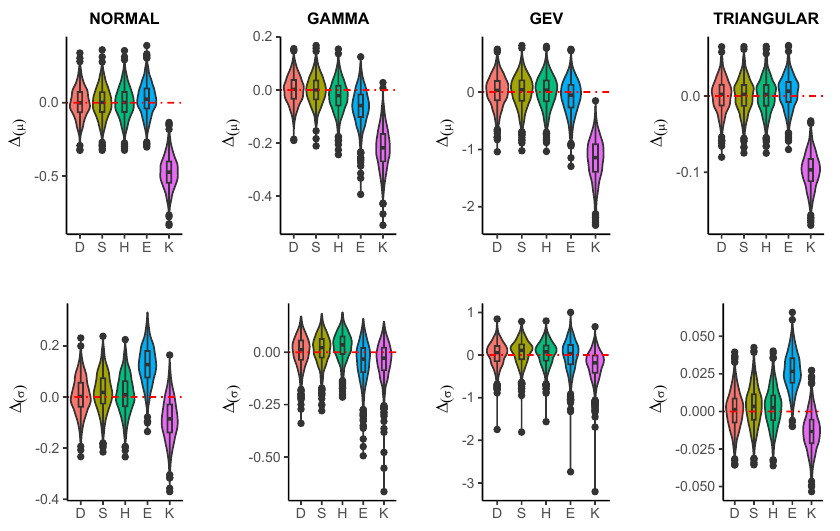}
		\captionsetup{font=scriptsize}
		\caption{\label{mean_SD_100} Pairwise differences of mean $\Delta(\mu)$ and standard deviation $\Delta(\sigma)$ estimates of binned data generated from an underlying continuous data with sample size $n=100$. D denotes estimates from the sampling distribution, and S, H, E, and K are estimates from the MICS, heuristic, eCDF, and kernel estimators, respectively. Dotted horizontal red lines represent the mark of over and underestimation of the underlying measure.}
	\end{figure}

	The calculated pairwise differences of the various metrics for the different distributions are shown in Figures~\ref{quantile_1000} and \ref{mean_SD_1000}, where D, for instance, denotes the pairwise difference of estimates from the sampling distribution and the theoretical values. Similarly, S, H, E, and K denote the pairwise differences of estimates from the MICS, heuristic, eCDF, and kernel estimators, respectively. Dotted red lines in the plots indicate if the different methods tend to overestimate or underestimate the actual values.
	
	As observed in Figure~\ref{quantile_1000}, in addition to approximating the actual theoretical quantities, the estimated quantiles with the MICS approach (S) outperform those from the popular and widely used approaches. They are also very close to those from the empirical distribution (D). It adapts very well to both mild and strong departures from normality (captured by the different distributions). In all four settings, the MICS approach further recovers much information about the spread of the underlying data that was lost due to binning.
	
	Similar patterns were also observed with the mean and standard deviation estimates reported in Figure~\ref{mean_SD_1000}. Besides yielding mean and standard deviation values that well approximate those from both the theoretical and empirical distributions (D), it stays consistent irrespective of changes in the shapes of the underlying distributions. The performance of the estimators was further assessed using a relatively small sample size $(n=100)$. Results are given in Figures~\ref{quantile_100} and \ref{mean_SD_100}, where we see almost a replica of the previous results, except for larger deviations from the theoretical values due to sampling errors. Meaning, that the MICS approach also performs very well, compared to competing methods, for smaller sample sizes. In general, the MICS's improvement over the heuristic and the other approaches is more pronounced with skewed distributions.

	\section{Real World Data Applications}\label{realDt}
	%-------------------------------------------------------------- 
	\subsection{Train Punctuality}\label{realDt1}
	%--------------------------------------------------------------
	For a real-world application of the discussed approaches, we consider the data on Deutsche Bahn (DB) delays \citep{deutschebahn_2022}. In the year 2022, only around two-thirds of all long-distance trains (ICE and IC) reportedly reached their destination on time (65.2\%). This has been attributed to poor and scarce infrastructure, intensive construction activity, and (from the second quarter) a rapidly growing traffic volume in long-distance and local transport, which all put rail operations under pressure. According to official records, long-distance transport went 10\% points below the previous year (75.2\%) in punctuality. Table~\ref{TrainPnct} shows the proportion of punctual stops in relation to all en route and final stops. A stop is considered on time by Deutsche Bahn if the arrival time is less than 6 or 16 minutes behind schedule, respectively. The corresponding aggregated data in terms of relative frequencies are published by DB on a monthly basis \citep{deutschebahn_2022}. The punctuality statistics in Table~\ref{TrainPnct} were obtained from more than 800,000 journeys by DB passenger trains (PT) in a month, all stops of more than 20,000 monthly trips in long-distance traffic (LT), and approximately 780,000 monthly trips in local traffic (RT) -- including all S-Bahn trains.

	\begin{table}[t]
		\captionsetup{font=scriptsize}
		\centering
		\caption{\label{TrainPnct} Aggregated delays of all DB passenger trains (PT), long-distance trains (LT), and regional trains (RT) as published by Deutsche Bahn; \% counts displayed.}
		\tabcolsep=6.1pt
		
		\begin{tabular}{ccccccc}
			\hline
			\hline
			& \multicolumn{6}{c}{\textbf{Deutsche Bahn Delays in 2022}}\\
			\cmidrule(lr){2-7} 
			& \multicolumn{2}{c}{\textbf{PT}}& \multicolumn{2}{c}{\textbf{LT}}& \multicolumn{2}{c}{\textbf{RT}} \\
			\cmidrule(lr){2-3} \cmidrule(lr){4-5} \cmidrule(lr){6-7}
			Month & $\le$ 5:59 Min & $\le$ 15:59 Min & $\le$ 5:59 Min & $\le$ 15:59 Min & $\le$ 5:59 Min & $\le$ 15:59 Min \\
			\hline
			Jan & 95.1 & 99.0 & 80.9 & 92.2 & 95.5 & 99.2 \\
			Feb & 93.9 & 98.6 & 76.0 & 88.7 & 94.4 & 98.8 \\
			Mar & 93.3 & 98.5 & 71.1 & 85.9 & 93.9 & 98.8 \\
			Apr & 93.2 & 98.5 & 69.1 & 85.2 & 93.9 & 98.8 \\
			May & 91.5 & 97.8 & 62.7 & 79.4 & 92.3 & 98.3 \\
			Jun & 87.7 & 96.7 & 58.0 & 77.3 & 88.5 & 97.3 \\
			Jul & 88.7 & 97.2 & 59.9 & 79.1 & 89.6 & 97.7 \\
			Aug & 88.6 & 97.1 & 56.8 & 76.8 & 89.5 & 97.7 \\
			Sep & 90.6 & 97.8 & 62.8 & 81.7 & 91.4 & 98.3 \\
			Oct & 89.8 & 97.7 & 63.2 & 82.0 & 90.6 & 98.1 \\
			Nov & 89.5 & 97.7 & 61.1 & 80.6 & 90.3 & 98.1 \\
			Dec & 89.8 & 97.5 & 61.3 & 79.6 & 90.7 & 98.0 \\
			\hline
		\end{tabular}
	\end{table}

	\begin{figure}[htb]\centering
		\includegraphics[width=150mm]{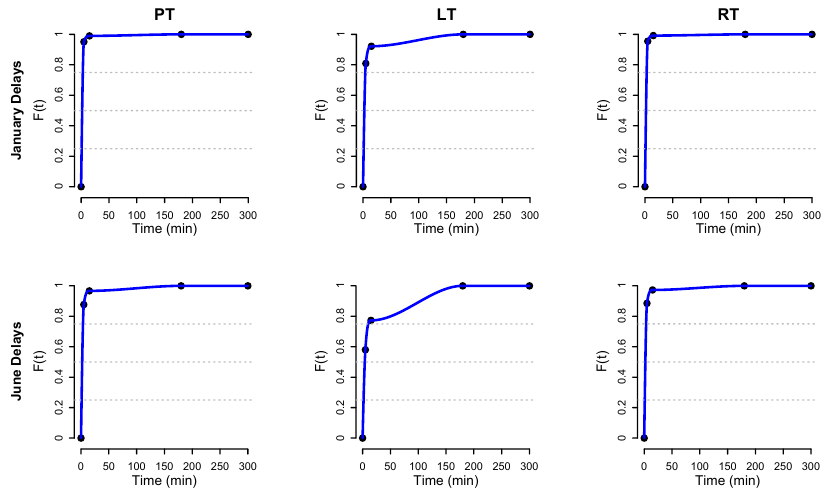}
		\captionsetup{font=scriptsize}
		\caption{\label{DB_splineCurves} Estimated cumulative distributions of the proportion of punctual stops in relation to all en route and final stops of the three classes of Deutsche Bahn (PT, LT, and RT) for the months of January and June.}
		\vspace{1cm}
	\end{figure}
	
	\begin{figure}[htb]\centering
		\includegraphics[width=120mm]{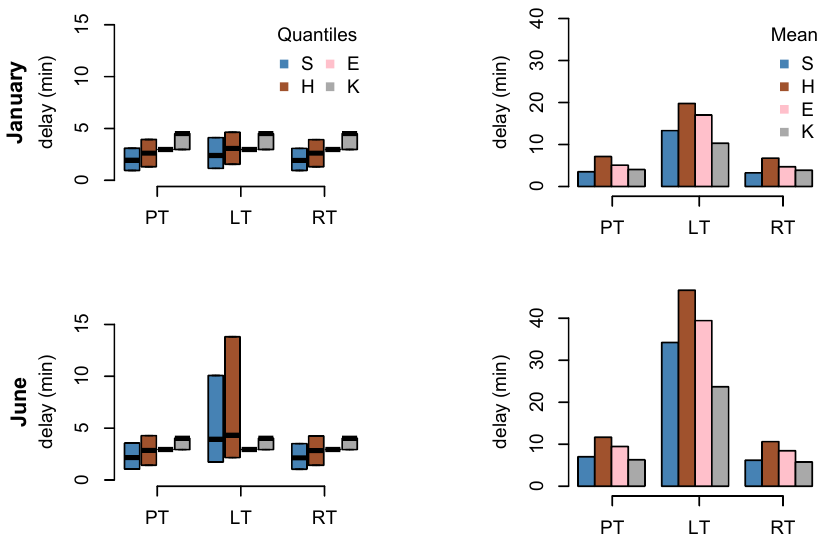}
		\captionsetup{font=scriptsize}
		\caption{\label{JanJuneDelays} Estimated population characteristics of delays in the months of January and June of Deutsche Bahn passenger trains (PT), long-distance trains (LT), and Regional trains (RT). Where S, H, E, and K are spline, heuristic, eCDF, and kernel estimates respectively.}
	\end{figure}
	
	\begin{table}[htb] %[ht]
		\captionsetup{font=scriptsize}
		\centering
		\caption{\label{AvgEstimate} Estimated population characteristics of the monthly and averaged delays of Deutsche Bahn long-distance trains (LT). The interquartile range (IQR) and the mean from the MICS (S), heuristic (H), eCDF (E), and kernel (K) estimators are reported.} 
		\tabcolsep=6.1pt
		
		\begin{tabular}{ccccccccc}
			\hline
			\hline
			& \multicolumn{2}{c}{\textbf{Spline}}& \multicolumn{2}{c}{\textbf{Heuristic}} & \multicolumn{2}{c}{\textbf{eCDF}}& \multicolumn{2}{c}{\textbf{Kernel}} \\ 
			
			\cmidrule(lr){2-3} \cmidrule(lr){4-5} \cmidrule(lr){6-7} \cmidrule(lr){8-9}
			
			Month & IQR& $\mu$& IQR  & $\mu$&  IQR& $\mu$&   IQR& $\mu$ \\
			\hline
			Jan   &2.97& 13.31&  3.09& 19.78& 0.00& 17.02&  2.63& 10.31 \\
			Feb   &3.54& 18.19&  3.29& 26.05& 0.00& 22.62&  0.44& 13.43 \\
			March &4.62& 22.15&  5.88& 31.16& 5.07& 27.01&  0.22& 15.96 \\
			April &5.02& 23.19&  6.86& 32.51& 5.08& 28.13&  5.04& 16.64 \\
			May   &7.03& 31.16& 10.37& 42.72& 5.08& 36.43&  5.04& 21.73 \\
			June  &8.34& 34.20& 11.65& 46.66& 5.10& 39.42& 32.03& 23.68 \\
			July  &7.35& 31.73& 10.78& 43.49& 5.10& 36.99&  5.05& 22.11 \\
			Aug   &8.73& 34.93& 11.90& 47.60& 5.10& 40.13& 62.10& 24.15 \\
			Sep   &6.35& 28.16&  9.46& 38.92& 5.09& 33.36&  5.04& 19.82 \\
			Oct   &6.24& 27.74&  9.30& 38.38& 5.09& 32.93&  5.04& 19.56 \\
			Nov   &6.77& 29.70& 10.08& 40.90& 5.10& 34.94&  5.04& 20.81 \\
			Dec   &7.07& 30.99& 10.45& 42.53& 5.09& 36.25&  5.04& 21.63 \\
			\hline
			Average &6.17& 27.12&  8.59& 37.56& 4.24& 32.10& 11.06& 19.15 \\
			\hline
		\end{tabular}
		\vspace{1cm}
	\end{table}

	Given the first month (January), for instance, it is seen that 95.1\% and 99.0\% of all passenger traffic trains respectively showed up not more than 5 and 15 minutes after schedule. The long-distance trains had 80.9\% and 92.2\%, while the regional trains had 95.5\% and 99.2\% respectively. However, with just two pairs of data points each not much could be said about the actual distribution of the trains' punctuality. For a better estimate of the underlying distribution, one at least considers two more extreme cases (data points), namely, 0\% and 100\% delays. The resulting four pairs of points could then be viewed as real points on the unobserved CDFs of train punctuality and could help to retrace the shape of the underlying distribution. While it makes sense in this scenario to assume that 0\% delays correspond to 0 minutes, one problem is how to fix the upper limit, $\tau_r$, at which delays correspond to 100\%. Per experience, delays may take up to 3 hours (180 mins). We would consider this the train delays' upper limit for estimating the underlying train delay CDFs. The obtained CDFs (via MICS) for the months of January and June for each of the three train types (PT, LT, and RT) are shown in Figure~\ref{DB_splineCurves}, with the last points at (300, 1) an additional pseudo observation used to stabilize the curve at the endpoints. On the one hand, the shapes of the CDFs of PT and RT look quite the same in January and June, suggesting a similar delay time frame. On the other hand, some differences are seen with the CDFs of the long-distance trains (LT), and, particularly for June delays, the estimated CDF looks odd. Interpolating straight lines would (presumably) produce more realistic results here.
	
	\begin{figure}[htb]\centering
		\includegraphics[width=120mm]{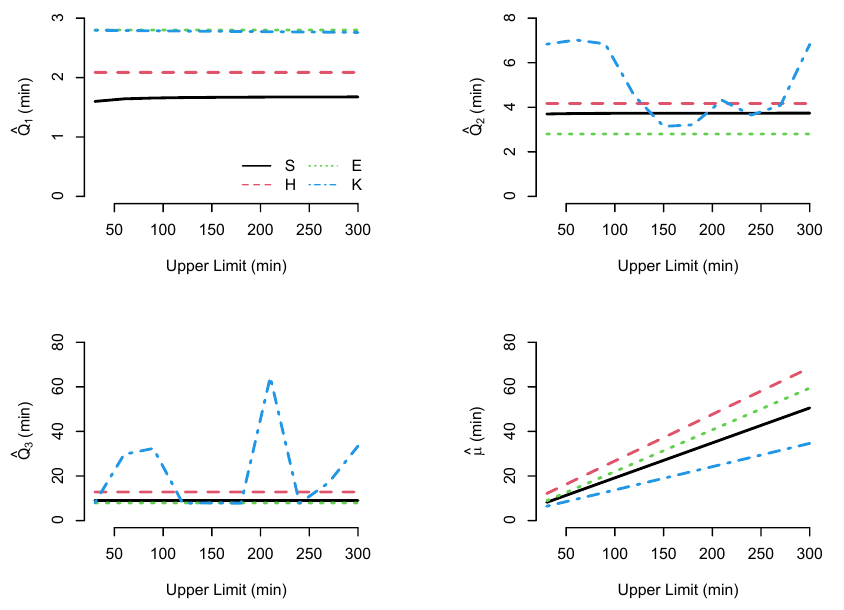}
		\captionsetup{font=scriptsize}
		\caption{\label{varyingLimit} Estimated population characteristics of delays of Deutsche Bahn long-distance train (LT) in the month of July using different topmost upper limits ($\tau_r$). Where S, H, E, and K are MICS, heuristic, eCDF, and kernel estimates respectively.}
	\end{figure}

	Also, while we could consider $\tau_r=180$ mins as the train delays upper limit, it could also be interesting to investigate what difference it makes using other upper limits on the different estimation methods. This is a well-known problem in several practical applications of binned data \citep[see, e.g.,][]{Jargowsky2018, Ho2012, Atkinson2011}. Thus, we investigate the effect of varying $\tau_r$ on the estimated population characteristics and consider the different estimation approaches. For this, we obtained the quantiles ($Q1$, $Q2$ and $Q3$) and means $(\mu)$ for the long-distance train (LT) given $\tau_r = \{30, \ 60, \ 120, \ 180 \}$. Results are given in Figure~\ref{varyingLimit}. As observed, quantile estimates (Q1, Q2, and Q3) from MICS (S), heuristic (H), and eCDF (E) approaches remain invariant to increasing $\tau_r$. On the other hand, the kernel estimates exhibited large variations with the median and third quartile, as indicated by up and down movements in the plots. The estimated means, on the other hand, increase monotonically with each rise in $\tau_r$. This is not surprising considering that even under normal circumstances, means are generally known to be affected by extreme values. In this instance, making decisions based on quantile estimates would be a good idea, rather than relying on the means.
	
	Continuing further with our fixed $\tau_r$ setting, the means and IQRs from the different approaches were subsequently obtained and compared in Figure~\ref{JanJuneDelays}, while Table~\ref{AvgEstimate} presents both the monthly and averaged estimates for the long-distance train (LT). Results show some form of dissimilarities between the different methods. As with some other estimates, the MICS estimates capture a realistic situation across the months, with values rising from January, climaxing in June till August (summer), dropping afterward, and picking up again in December. Furthermore, estimates from E and K are smaller than those from S and H, mainly due to less variability captured by the former, particularly with E which is basically a step function with constant values recorded at different quantile steps. The long-distance train (LT) out of the three train categories had the largest estimated mean delays and widest spread (in terms of IQR). On average (i.e., annually), the estimated mean delays in minutes and IQR in parenthesis for 2022 using MICS are 27.12 (6.17); 37.56 (8.59), 32.10 (4.24), and 19.15 (11.06) for the heuristic, eCDF and kernel estimators, respectively.

	%-------------------------------------------------------------- 
	\subsection{Distance-Time to Work}\label{realDt2}
	%--------------------------------------------------------------
	\begin{table}[htb]
		\captionsetup{font=scriptsize}
		\centering
		\caption{\label{DisTime} Upper segment: \% frequencies of the distance in kilometers traveled by all workers (All), self-employed workers (SE), and the employees (EM) to work in Germany. Lower segment: the amount of time in minutes spent en route to the workplace. Actual observations are given in groups with the lower (LL) and upper class limits (UL) also indicated.} 
		\tabcolsep=6.1pt
		
		\begin{tabular}{lrrrr}
			\hline
			&&  \multicolumn{3}{c}{\textbf{Commuters}} \\
			\cmidrule(lr){3-5} 
			LL&    UL&    All-workers (\%)&      Self-employed (\%)&        Employees (\%)\\
			\hline
			\multicolumn{2}{l}{\textbf{Distance (km)}}& && \\ 
			
			\cmidrule(lr){1-2}
			
			0&     5&   27.5&     47.7&   26.3 \\
			5&    10&   22.6&     20.0&   22.7 \\
			10&    25&  30.1&     20.4&   30.8\\
			25&    50&  14.6&      7.7&   15.1\\
			50&   100&   5.2&      4.2&    5.2\\ 
			
			\hline
			\hline
			\multicolumn{5}{l}{\textbf{Time (min)}}  \\
			
			\cmidrule(lr){1-2}
			
			0&    10 &  21.4&     41.6&   20.2 \\
			10&    30&   50.9&     41.7&   51.5\\
			30&    60&   22.7&     13.1&   23.2\\
			60&    90&    5.0&      3.7&    5.1\\
			
			\hline
		\end{tabular}
		\vspace{1cm}
	\end{table}

	\begin{figure}[htb]\centering
		\includegraphics[width=150mm]{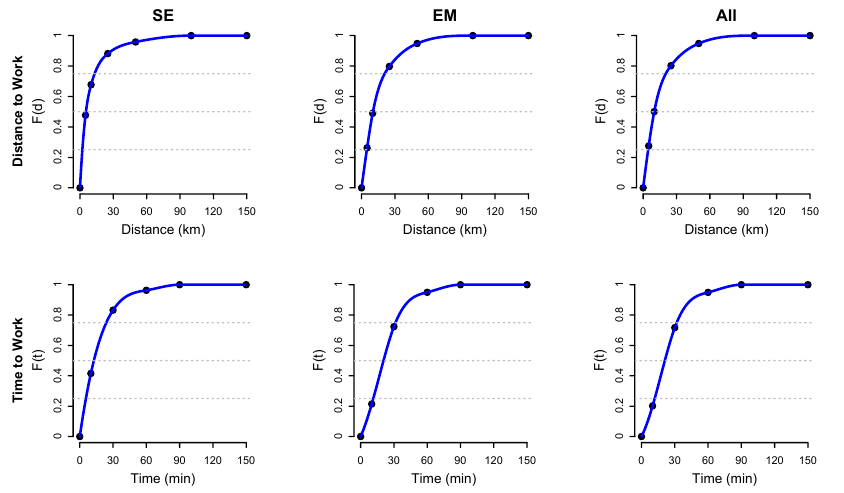}
		\captionsetup{font=scriptsize}
		\caption{\label{splinCuv} Estimated cumulative distributions of distance in kilometers to work and time in minutes to work of the self-employed (SE), employees (EM), and all workers (All) in Germany.}
	\end{figure}

	The next real data example deals with a segment of micro-census data on the distance and time required to get to work by workers in Germany in 2020. According to \cite{statistisches_bundesamt_2022}, the term `micro-census' means ``small population census'' and represents the largest annual household survey of official statistics in Germany. The survey has been carried out jointly by the federal and state statistical offices since 1957. Hence, with roughly 810,000 people in around 370,000 private households and communal accommodations, about 1\% of Germany's population was surveyed about their working and living conditions. The statistics office had the federal territory divided into areas with approximately the same number of apartments, giving all districts/cities equal chances of being selected for the survey. Thus, selection for the survey depends on a respondent's living apartment being selected. Table 4 shows the \% frequencies of distance in kilometers and the time in minutes traveled to work by all workers (All), the self-employed (SE), and the employees (EM).
	
\begin{figure}[htb]\centering
		\includegraphics[width=120mm]{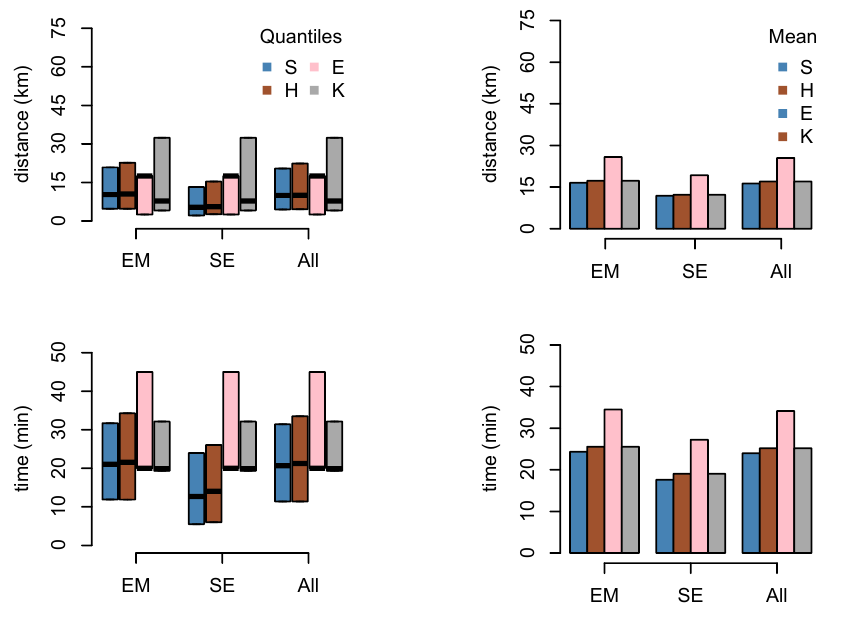}
		\captionsetup{font=scriptsize}
		\caption{\label{disTimeQtl} Estimated population characteristics of the distance in kilometers (top row) and time in minutes (bottom row) of employees (EM), self-employed (SE), and all workers (All) in Germany.  Where S, H, E, and K are spline, heuristic, eCDF, and kernel estimates, respectively.}
	\end{figure}

	Similar to the previous example, the distribution of distance and time to work for all workers, the self-employed, and the employees were estimated using MICS and the other estimation methods. The estimated CDFs are shown in Figure~\ref{splinCuv}, with the SE curve steeper than EM in both distance and time CDFs. The means, first, second, and third quartiles from the estimated distribution were obtained using the MICS and other estimation methods (see Figure~\ref{disTimeQtl}). While MICS and heuristic results do not differ so much, both differ from the other two approaches and are more and more pronounced with the quantile estimates. With the exception of the estimated mean from the eCDF, the estimated mean from the other three is nearly uniformly shaped. In summary, the obtained results suggest that employees in Germany spend more time and distance to work than self-employed individuals.
	
	\section{Discussion}\label{disc}
	
	Nonparametric density estimation provides an alternative means of obtaining crucial estimates of different population characteristics in situations where the parametric approach fails. In this study, we have discussed how such could be used to estimate the distribution of binned data. Rather than handling binned data in an aggregated form, leading to the loss of valuable information, we would rather estimate the underlying (unobserved) CDF with monotone-preserving cubic splines, often leading to more accurate and reliable estimates. This approach frequently outperforms other widely used approaches, as shown in our simulation studies and real data examples.
	
	However, as observed in our study, a typical open question while estimating the population characteristics of binned data is how to handle open-ended classes, a common problem in various large-scale surveys. In situations with no standard benchmark limits to fall on, this could present a daunting challenge to making reliable inferences from binned data. As noted earlier in Section~\ref{intro}, common suggestions, such as those in studies of income distributions, include assuming that the quantity in the upper tail of the distribution follows a Pareto distribution \citep{hippel2015a}, which we consider a costly assumption. \cite{Jargowsky2018} proposed an improvement on this approach by introducing the mean-constrained integration over brackets (MCIB) approach, which retains the same Pareto distribution assumption for the upper tail but employs a different approach for estimating the shape parameter of the specified distribution. Alternatively, \cite{Rizzi2015} demonstrated how binned data can be ungrouped using the penalized composite link model. This approach can handle wide, open-ended intervals and incorporate extra information about how the tail area could be occupied, assuming such information is available. However, estimation with this approach is based on an iteratively reweighted least-squares algorithm, with the selection of a smoothing parameter through the minimization of AIC, which is inherently parametric. Another proposal by \cite{vhippel2017a} involves constraining the parameter shaping the top bin to the grand mean if known, such that the mean of the fitted distribution matches the grand mean. When the grand mean is unknown, an ad hoc estimate is obtained by temporarily setting the upper bound in question to twice the lower bound, calculating the mean of a step PDF fit to all bins, unbinding the top bin, and proceeding as though the mean were known.
	
	Notwithstanding these approaches, the inherent subjectivity associated with them leaves the initial problem not entirely solved. In our study, we investigated the effect of varying upper limits in our Deutsche Bahn delay study, for instance, and observed that some population characteristics are not invariant to the choice of missing $\tau_r$. The mean, in particular, is most affected. However, we did note that quantile estimates remain invariant to the choice of $\tau_r$ and should be reported in scenarios where the mean estimate could be influenced by extreme values.
	
	In conclusion, we consider the MICS density estimation presented in this study a promising non-parametric approach for handling binned data. The certainty of having the interpolating curve pass through all data points is assured, which is preferable to fitting inflexible curves that fail to capture the complex variation of actual data. Moreover, the MICS approach easily adapts to different underlying distributional shapes, as demonstrated in the simulation study, and is not affected by whether bin sizes are equally or unequally spaced.
	
	\clearpage
	\bibliographystyle{chicago}
	\bibliography{cdf}
\end{document}